\documentclass[aps,prd,superscriptaddress,showpacs,showkeys,nofootinbib,nobibnotes,preprintnumbers,floatfix]{revtex4}

\usepackage{amsmath}
\usepackage{amssymb}
\usepackage{graphicx}
\usepackage{color}
\usepackage{curves}
\usepackage{bm}  
\usepackage{slashed}
\usepackage{enumerate}

\newcommand{\alphir}{\alpha_{\mathrm{IR}}}
\newcommand{\chpt}{\chi\mbox{PT}}
\newcommand{\fpi}{F_{\pi}}
\newcommand{\fsig}{F_{\sigma}}
\newcommand{\tr}{\mathrm{Tr}}
\newcommand{\del}{\partial}
\newcommand{\rir}{R_{\mathrm{IR}}}

\begin{document}

\title{Infrared Fixed Point in the Strong Running Coupling: \\
Unraveling the $\Delta I=1/2$ puzzle in $K$-Decays%
\footnote{Contribution to the proceedings of the workshop 
``Determination of the Fundamental Parameters of QCD",
Nanyang Technological University, Singapore, 18-22 March 
2013, to be published in Mod.\ Phys.\ Lett. A.}}

\author{R.~J.~Crewther}
\email{rcrewthe@physics.adelaide.edu.au}
\affiliation{CSSM and ARC Centre of Excellence for Particle Physics at 
the Tera-scale, School of Chemistry and Physics, University of Adelaide, 
Adelaide SA 5005 Australia}
\author{Lewis~C.~Tunstall\footnote{Speaker.}}
\email{lewis.tunstall@adelaide.edu.au}
\affiliation{CSSM and ARC Centre of Excellence for Particle Physics at 
the Tera-scale, School of Chemistry and Physics, University of Adelaide, 
Adelaide SA 5005 Australia}

\begin{abstract}
In this talk, we present an explanation for the $\Delta I=1/2$ rule in $K$-decays based 
on the premise of an infrared fixed point $\alphir$ in the running coupling 
$\alpha_s$ of quantum chromodynamics (QCD) for three light quarks $u,d,s$.  
At the fixed point, the quark condensate $\langle\bar{q}q\rangle_{\mathrm{vac}} \neq 0$ 
spontaneously breaks scale {\it and} chiral $SU(3)_L\times SU(3)_R$ symmetry.  
Consequently, the low-lying spectrum contains {\it nine} Nambu-Goldstone bosons: 
$\pi,K,\eta$ and a QCD dilaton $\sigma$.  We identify $\sigma$ as the $f_0(500)$ 
resonance and construct a chiral-scale perturbation theory $\chpt_\sigma$ for 
low-energy amplitudes expanded in $\alpha_s$ about $\alphir$.  The 
$\Delta I=1/2$ rule emerges in the leading order of $\chpt_\sigma$ through a 
$\sigma$-pole term $K_S\to\sigma\to\pi\pi$, with a $g_{K_S\sigma}$ coupling 
fixed by data on $\gamma\gamma\to\pi^0\pi^0$ and $K_S\to\gamma\gamma$.  
We also determine $R_{\mathrm{IR}} \approx 5$ for the nonperturbative Drell-Yan 
ratio at $\alphir$.

\end{abstract}


\maketitle

\section{The $\Delta I =1/2$ Puzzle} 
Amidst the rich phenomenology of $K$-mesons lies a severe problem%
---so old that new solutions are rarely attempted---associated with the 
strangeness-changing $|\Delta S|$=1, nonleptonic decays of the 
short- and long-lived states
\begin{equation}
K_S\to\pi\pi\,, \qquad K_L\to\pi\pi\pi\,.
\end{equation}  
Experimentally, there is a large enhancement of the isospin-%
$\tfrac{1}{2}$ decays.  This phenomenon is particularly striking in 
the $S$-wave $\pi\pi$ mode, where the measured rates \cite{PDG} 
exhibit the ratios
\begin{equation}
\gamma_{+-} = \frac{\Gamma(K_S\to\pi^+\pi^-)}{\Gamma(K^+\to\pi^+\pi^0)} 
\simeq 463\,, \qquad 
\gamma_{00} = \frac{\Gamma(K_S\to\pi^0\pi^0)}{\Gamma(K^+\to\pi^+\pi^0)} 
\simeq 205\,,
\label{eqn:rates}
\end{equation}
which are in enormous disagreement with the naive expectations 
$\gamma_{+-} \sim \mbox{O}(1) \sim \gamma_{00}$ from perturbative 
electroweak calculations.  It is useful to translate the above in terms of 
isospin amplitudes $A_I$ for the final $\pi\pi$ state.  The $I=1$ state is forbidden 
by Bose symmetry and thus the transition amplitudes can be 
parametrized as \cite{Mar69}
\begin{align}
{\cal A}(K_S\to \pi^+\pi^-) &= \frac{2}{\sqrt{3}} A_0e^{i\delta_0} 
+ \sqrt{\frac{2}{3}} A_2e^{i\delta_2}\,, \\
{\cal A}(K_S\to\pi^0\pi^0) &= \sqrt{\frac{2}{3}} A_0e^{i\delta_0} 
- \frac{2}{\sqrt{3}} A_2e^{i\delta_2}\,, \\
{\cal A}(K^+\to\pi^+\pi^0) &= \sqrt{\frac{3}{2}} A_2e^{i\delta_2}\,,
\end{align}
where the $\pi\pi$-scattering phase shifts $\delta_I$ arise as a 
consequence of Watson's theorem.  Comparison with 
the data in (\ref{eqn:rates}) leads to the 
$\Delta I=1/2$ rule for kaons\footnote{A similar rule is observed in the nonleptonic hyperon decays.}
\begin{equation}
\Re\mathfrak{e} |A_0/A_2| \simeq 22\,,
\label{eqn:ratio}
\end{equation}
whose origin remains a mystery despite five decades of theoretical 
investigation.

The difficulty presumably arises from the nonperturbative nature 
of quantum chromodynamics (QCD) at low energies $\mu \ll m_{t,b,c}$, 
where confinement reigns and chiral symmetry for the light quarks 
$u,d,s$ is believed to be spontaneously broken by the formation of a 
quark condensate
\begin{equation}
\lim_{m_q\to 0} \langle\bar{q}q\rangle_{\mathrm{vac}}\neq 0\,, 
\qquad q = u,d,s\,.
\label{eqn:chiral limit}
\end{equation}
Below the chiral symmetry-breaking scale $\Lambda_{\chi\mathrm{SB}} \approx 1$ 
GeV, great progress has been made in charting the low-energy structure of QCD 
through the use of chiral $SU(3)_L \times SU(3)_R$ perturbation theory $\chpt_3$.  
The method relies on an alternative expansion parameter to the strong coupling 
$\alpha_s = g^2/4\pi$, {\it viz}.\ low-energy scattering amplitudes and matrix 
elements can be described by an asymptotic series
\begin{equation}
{\cal A} = \{ {\cal A}_{\mathrm{LO}} + {\cal A}_{\mathrm{NLO}} + \ldots \}
\label{eqn:asy exp}
\end{equation}
in powers of $O(m_K)$ momentum and quark masses $m_{u,d,s} = O(m_K^2)$.  
The amplitudes of the series are then calculated in a systematic manner by constructing 
the most general effective Lagrangian consistent with the underlying symmetries 
of QCD.  For example, the strong interactions of $\pi,K,\eta$ mesons are 
described in the leading order (LO) of $\chpt_3$ by 
\begin{equation}
\left.{\cal L}_{\mathrm{str}}\right|_{\mathrm{LO}} = \frac{\fpi^2}{4}\tr(\del_\mu U\del^\mu U^\dagger) 
+ \tr(MU^\dagger + UM^\dagger)\,,
\end{equation}
where $F_\pi\simeq 93$ MeV is the pion decay constant, and $M$ is proportional 
to the $u,d,s$ quark mass matrix.  Here $U=U(\phi)$ is an $SU(3)$ field, where 
the octet of pseudoscalar Nambu-Goldstone (NG) bosons $\phi_{i}$ parametrize 
the coset space $(SU(3)_{L}\times SU(3)_{R})/SU(3)_{V}$ with group action
\begin{equation}
U \rightarrow RUL^{\dagger}\,, \qquad R\in SU(3)_{R}\,,\qquad L\in SU(3)_{L}\,.
\end{equation}

The method can likewise be applied to the weak interactions and used to simulate 
nonleptonic decays.  The chiral structure of the weak currents $(8_L,1_R)\oplus (27_L,1_R)$ 
constrains the number of allowed operators.  For instance, in the leading order of 
$\chi$PT$_3$, two octets can be constructed with the required transformation 
properties, {\it viz}.\ the derivative operator \cite{Cro67}
\begin{equation}
Q_{8} = \mathcal{J}_{13}\mathcal{J}_{21} - \mathcal{J}_{23}\mathcal{J}_{11}
\ , \quad
\mathcal{J}_{ij} = (U\partial_{\mu}U^{\dagger})_{ij}
\end{equation}
and a weak mass operator \cite{Bern85}
\begin{equation}
Q_M = \mathrm{Tr} (\lambda_6 - i\lambda_7)
      \bigl(g_MMU^\dagger + \bar{g}_MUM^\dagger\bigr) \,.
\end{equation}
As an $SU(3)$ octet, $Q_M$ is necessarily isospin-$\tfrac{1}{2}$, however 
the requirement of a stable vacuum ensures that this operator does not 
contribute to physical processes.  In this regard, it can be shown \cite{RJC86} 
that the effective potential is minimized by a chiral rotation such that 
$\langle U\rangle_{\mathrm{vac}} = I$ and $M$ is diagonal and positive.  
By aligning the vacuum in this way, all trace of $Q_M$ is removed from the 
effective Lagrangian, and thus by including the $U$-spin triplet 
component \cite{RJC86,Gaill74} of a \textbf{27} operator
\begin{equation}
Q_{27} = \mathcal{J}_{13}\mathcal{J}_{21} 
           + \tfrac{3}{2} \mathcal{J}_{23}\mathcal{J}_{11}\,,
\end{equation}
the nonleptonic decays are described by the compact expression
\begin{equation}
\mathcal{L}_{\mathrm{weak}}
= g_{8}Q_{8} + g_{27}Q_{27} + \mathrm{h.c.}\,.\ 
\label{usual}\end{equation}

In general, the low-energy coefficients $g_i$ are not fixed by symmetry 
arguments alone.  This constitutes the essence of the $\Delta I=1/2$ puzzle:  
why is $|g_8/g_{27}|$ unreasonably large ($\approx$ 22) when compared 
with simple quark-model estimates for $\Delta I=3/2$ $K$-decay 
amplitudes?  Explaining this `octet dominance' has been tackled with a variety 
of techniques including the many color $N_c$ limit \cite{Tad82,Bur86,Chiv86,Bar86,Bij88}, 
QCD sum rules \cite{Gub85,Pich86,Pich87,Pich91}, and direct evaluation on the 
lattice \cite{Blum11,Blum12,Boy13}, each with varying degrees of success.

\section{Infrared Fixed Point in the Strong Running Coupling}
The goal of this talk is to present an alternative line of investigation \cite{CT12} and 
argue that the $\Delta I=1/2$ rule for $K$-decays (Eq.~(\ref{eqn:ratio})) 
is intimately linked to the behavior of $\alpha_s$ in the {\it asymptotic} infrared 
limit.  As is well known, the running of $\alpha_s$ with scale $\mu$ is 
governed by the QCD $\beta$-function. At low energies $\mu \ll m_{t,c,b}$, 
heavy quarks $t,b,c$ decouple from the theory, and thus there are two 
logical possibilities\footnote{The analogous case for QED is discussed in 
Ref.~\cite{Weinberg}.} for the resulting three-flavor theory (Fig.~\ref{fig:beta}):
%
%%% Fig.~\ref{fig:beta}
\begin{figure}[t]
\center\includegraphics[scale=.7]{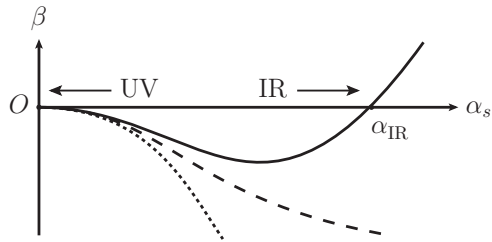}
\caption{Varieties of asymptotic behavior for the QCD $\beta$-function 
with three light quarks $u,d,s$.  The dashed line shows the case when 
the strong running coupling $\alpha_s$ undergoes continued growth 
with decreasing scale $\mu$ (item (\ref{unbound}) in text), while the solid 
line shows $\alpha_s$ flowing to an infrared fixed point $\alphir$ (item 
(\ref{ir fixed pt})).  For completeness, we also include the case where 
$\alpha_s$ diverges at a finite value of $\mu$ (dotted line), but we emphasize 
that this scenario is of {\it no physical relevance} since it produces poles 
in Green's functions in the spacelike momentum region (i.e.\ a tachyon 
or ``Landau ghost").
\label{fig:beta}}
\end{figure}
\begin{enumerate}[i]
\item {\bf Growth without bound.}  If the integral
\begin{equation}
\int_{\alpha_0}^\infty \frac{dz}{\beta(z)}
\label{eqn:rge int}
\end{equation}
is divergent, then the solution to the renormalization group equation
\begin{equation}
\ln\mu/\mu_0 = \int_{\alpha_0}^{\alpha_s} 
\frac{dz}{\beta(z)}
\end{equation}   
implies that as $\mu$ decreases, $\alpha_s$ grows in value, becoming infinite 
in the infrared limit $\ln\mu/\mu_0\to -\infty$.  Analyses based on lattice 
simulations \cite{Lusc94} suggest that this scenario occurs for $N_f=0$ quark 
flavors (pure Yang-Mills).\label{unbound}

\item {\bf Infrared fixed point at finite coupling.}  In this scenario, 
\begin{equation}
\int_{\alpha_0}^{\alphir} \frac{dz}{\beta(z)}
\end{equation}
diverges because of a simple zero in $\beta(z)$ at an infrared fixed point $z = \alphir$.  
From a purely phenomenological point of view, this situation is perhaps the most 
interesting for it implies that the gluonic term $\sim G_{\mu\nu}^a G^{a\mu\nu}$ 
in the trace anomaly of the energy-momentum tensor $\theta_{\mu\nu}$,
\begin{equation}
\theta^\mu_\mu =\frac{\beta(\alpha_{s})}{4\alpha_{s}} G^a_{\mu\nu}G^{a\mu\nu}
 + \bigl(1 + \gamma_{m}(\alpha_{s})\bigr)\sum_{q=u,d,s} m_{q}\bar{q}q
\label{eqn:anomaly}
\end{equation}
vanishes at the fixed point $\alphir$ and thus $\theta_{\mu\nu}$ becomes traceless 
in the chiral limit
\begin{align}  
\left.\theta^\mu_\mu\right|_{\alpha_s = \alphir}
 &= \bigl(1 + \gamma_{m}(\alphir)\bigr)
    (m_u\bar{u}u + m_d\bar{d}d + m_s\bar{s}s) \nonumber \\
 &\to 0\ , \, SU(3)_L\times SU(3)_R \mbox{limit}\,.
\label{scale}
\end{align}
Naively, one might consider this scenario a phenomenological disaster: does 
a traceless $\theta_\mu^\mu$ not imply invariance under scaling transformations 
$\xi:x\to e^\xi x$ and hence a continuous mass spectrum?  In QCD, the answer 
is in the negative since the attendant strong gluon fields still drive quark condensation 
$\langle\bar{q}q\rangle_{\mathrm{vac}}\neq 0$, from which one concludes that 
the scale---and of course, chiral---symmetry is in fact {\it spontaneously broken}.%
\footnote{The idea that $\langle\bar{q}q\rangle_{\mathrm{vac}}$ may act as a 
chiral-scale condensate was considered prior to the advent of QCD in 
Refs.~\cite{RJC70,Ell70}.} Goldstone's theorem and approximate chiral-scale symmetry 
then implies that there are nine pseudo-NG bosons in the low-lying spectrum: the 
usual pseudoscalar octet $\pi,K,\eta$ and a $0^{++}$ QCD dilaton $\sigma$, 
whose mass is (mostly) set by the explicit breaking term $\propto m_s$ in 
(\ref{scale}).\label{ir fixed pt}
\end{enumerate}

It is worth emphasizing that it is unclear from the literature which scenario 
is actually realized in QCD, and in particular, how sensitive the results are 
to the number of active quark flavors.  Part of the problem resides in the 
fact that for each calculation, a nonperturbative\footnote{Extrapolations 
based on fixed-order perturbation theory are likely to introduce non-physical 
artifacts.} {\it definition} for $\alpha_s$ must be chosen, and thus comparing 
results from different approaches is rarely straightforward.  Popular choices 
include the Schr\"{o}dinger functional method on the lattice, \cite{Lusc92,Aoki09} 
solutions to Dyson-Schwinger equations \cite{vonSme97,Fisc03} for three-gluon 
and ghost- or  quark-gluon vertices, the method of effective charges \cite{Grun80,Grun84}, 
and AdS/CFT inspired models \cite{Brod10}.

Despite the lack of consensus, we take the view that an infrared fixed point in QCD 
(item (\ref{ir fixed pt})) should be taken seriously, and the remainder of this talk concerns the 
phenomenological implications which arise from this scenario.  

\section{The Lowest QCD Resonance as a QCD Dilaton}
While it may seem dramatic to introduce a new NG boson into the low-energy 
spectrum of QCD, there is in fact a very natural candidate for $\sigma$: the 
$f_0(500)$ resonance, whose mass and width have been determined with 
remarkable precision through an analytic continuation of the Roy equations \cite{Cap06},
\begin{equation}
M_{f_0} = 441^{+16}_{-8}\,\mathrm{MeV}\,, \quad 
\Gamma_{f_0} = 544^{+18}_{-25}\,\mathrm{MeV}, \quad 
|g_{f_0\pi\pi}| = 3.31^{+0.35}_{-0.15}\,\mathrm{GeV}\,.
\label{eqn:f0 pole}
\end{equation}
Since this pioneering work, many studies (see e.g. Fig.~8 in Ref.~\cite{Alba12}) 
have extracted $f_0$ pole parameters consistent with (\ref{eqn:f0 pole}), to the 
extent that the Particle Data Group \cite{PDG} has recently updated their 
listing to account for the greatly reduced uncertainties.\footnote{See 
Ref.~\cite{Pel13} for a discussion on this point.}  

\subsection{The $f_0(500)$ and $\chpt_3$: Problems with $SU(3)_L \times SU(3)_R$?}
In all determinations of the kind given by Ref.~\cite{Cap06}, the real part 
of the $f_0$ pole is found to be $\lesssim m_K$.  This places $f_0$ right in 
the middle of the NG sector $\pi,K,\eta$ and encourages us to revisit the 
observation \cite{Meiss91} that $\chpt_3$ works well except for $0^{++}$ 
amplitudes with $O(m_K)$ extrapolations in momentum.  Since $f_0$ is a 
broad flavor singlet coupled strongly to $\pi,K,\eta$, it should dominate 
low-energy scattering.  However, this leads to the following problem: $f_0$ 
cannot contribute to the LO terms ${\cal A}_{\mathrm{LO}}$ in (\ref{eqn:asy exp}) 
so the dominant effect must be generated at next to leading order (NLO).  
How does this square with the expectation that NLO terms are 
$\lesssim 30\%$ the LO prediction?  

One alternative (and the one we shall pursue) is to note that if $f_0=\sigma$ 
is treated a pseudo-NG boson (dilaton), then all the convergence issues in 
the $0^{++}$ channel disappear.  To that end, we replace $\chpt_3$ with 
a model-independent chiral-scale theory $\chpt_\sigma$ based on expansions 
in $\alpha_s$ about $\alphir$.

\section{Effective Lagrangians for Approximate Chiral-Scale Symmetry}
\subsection{Strong Interactions}
Our task is to construct an effective field theory of approximate scale and 
chiral $SU(3)_L\times SU(3)_R$ symmetry.  For the strong interactions we seek 
an effective Lagrangian of the form
\begin{equation}
{\cal L}[\sigma, U, U^\dagger] = :{\cal L}^{d=4}_{\mathrm{inv}} 
+ {\cal L}^{d>4}_{\mathrm{anom}} + {\cal L}^{d<4}_{\mathrm{mass}}:\,,
\end{equation}
where each term is distinguished by the scaling dimension $d$:
\begin{equation}
\delta_\xi {\cal L}_d = \del^\lambda (x_\lambda{\cal L}_d) + (d-4){\cal L}_d\,.
\end{equation}
The rules for constructing Lagrangians of this type were worked out long ago 
and we refer to Refs.~\cite{Ell70,Salam,IshamI,IshamII} for explicit 
details.  The salient features are as follows.  For expansions in $\alpha_s$ about 
$\alphir$, we have
\begin{equation}
d_{\mathrm{mass}} = 3 - \gamma_m(\alphir) < 4\,,
\end{equation}
while the gluonic operator insertion $\sim G_{\mu\nu}^aG^{a\mu\nu}$ (obtained 
by acting $\del/\del\alpha_s$ on the Callan-Symanzik equation) implies 
\begin{equation}
d_{\mathrm{anom}} = 4 + \beta'(\alphir) > 4\,.
\end{equation}

To realize scale invariance as a NG symmetry, a dilaton field $\sigma$ is introduced 
with nonlinear scaling property $\xi:\sigma\to\sigma + \xi\fsig$ so that $e^{\sigma/\fsig}$ 
is covariant and has $d=1$:
\begin{equation}
\delta_\xi e^{\sigma/\fsig} = (1 + x\cdot\del) e^{\sigma/\fsig}\,.
\end{equation}
Here $\fsig$ is the dilaton decay constant, defined by the coupling of 
$\theta_{\mu\nu}$ to the vacuum
\begin{equation}
\langle\sigma|\theta_{\mu\nu}|\mbox{vac}\rangle = 
\frac{\fsig}{3}(q_\mu q_\nu - g_{\mu\nu}q^2)\,.
\end{equation}
Lagrangian operators of the required scale dimension are then obtained 
by multiplying operators such as 
\begin{equation}
{\cal K} = \frac{\fpi^2}{4}\tr(\del_\mu U\del^\mu U^\dagger)\,, 
\qquad {\cal K}_\sigma = \frac{1}{2}\del_\mu\sigma\del^\mu\sigma\,,
\end{equation}
by appropriate powers of $e^{\sigma/\fsig}$.  In the LO of $\chpt_\sigma$ 
the most general effective Lagrangian for strongly interacting $\pi,K,\eta$ and $\sigma$ 
mesons is given by
\begin{align}
\mathcal{L}^{d=4}_\mathrm{inv}
 &= \bigl\{c_{1}\mathcal{K} + c_{2}\mathcal{K}_\sigma 
     + c_{3}e^{2\sigma/F_{\sigma}}\bigr\}e^{2\sigma/F_{\sigma}} 
\\
\mathcal{L}^{d>4}_\mathrm{anom} &= \bigl\{(1-c_{1})\mathcal{K} + (1-c_{2})\mathcal{K}_\sigma
      + c_4 e^{2\sigma/F_{\sigma}}\bigr\}e^{(2+\beta')\sigma/F_{\sigma}}
\\
\mathcal{L}^{d<4}_\mathrm{mass} 
 &= \mathrm{Tr}(MU^{\dagger}+UM^{\dagger})e^{(3-\gamma_m)\sigma/F_{\sigma}}\,. 
\end{align}
The low-energy constants $c_{i}$ are not fixed  by symmetry 
arguments, however, the requirement of a stable vacuum in the 
$\sigma$ direction implies that $c_{3}$ and $c_{4}$ are $O(M)$ and not 
independent. For expansions about $\sigma=0$, the absence of tadpoles 
implies that all terms linear in $\sigma$ must cancel:
\begin{align}
4c_3 + (4+\beta')c_4 &= 
(\gamma_m-3)\langle \mathrm{Tr}(M U^{\dagger}+UM^{\dagger}) \rangle_{\mathrm{vac}} \notag \\
&= (\gamma_m-3)F^{2}_{\pi}(m_{K}^{2} + \tfrac{1}{2}m^{2}_{\pi})\,.
\end{align}

\subsubsection{Determining $\fsig$}
\begin{figure}[t]
\center\includegraphics[scale=.7]{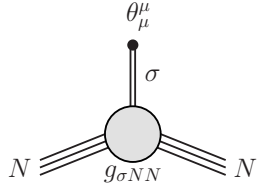}
\caption{Dominant $\sigma$-pole diagram in $\chi$PT$_\sigma$ for
$\langle N|\theta^\mu_\mu |N\rangle$.} 
\label{fig:g_sigNN}
\end{figure}

The relationship between $\fsig$ and $g_{\sigma NN}$ is deduced 
by an analogue of the Goldberger-Treiman relation (Fig.~\ref{fig:g_sigNN}):
\begin{equation}
\langle N|\theta^\mu_\mu|N\rangle = M_N 
= (-m_{\sigma}^2F_\sigma)\cdot(-i/m_\sigma^2)(-ig_{\sigma NN}) 
= F_\sigma g_{\sigma NN}\,.
\end{equation}
%
%%% Fig.~\ref{fig:g_sigNN}
One-boson exchange combined with large-$N_c$ arguments \cite{CC08,CC10} leads to 
the estimate $g_{\sigma NN} \simeq 9$ with a large model-dependent uncertainty. 
We take the central value and obtain $\fsig \simeq $ 100 MeV.

\subsubsection{Strong phenomenology: $\pi\pi$-scattering}
From ${\cal L}$ we obtain the dilaton mass
\begin{equation}
m_\sigma^2 F_\sigma^2 = 
F_\pi^2 (m_K^2 + \tfrac{1}{2}m_\pi^2)(3-\gamma_m)(1+\gamma_m) 
- \beta'(4+\beta')c_4\,, 
\end{equation}
and the effective $g_{\sigma\pi\pi}$ coupling
\begin{equation}
\mathcal{L}_{\sigma\pi\pi}
 = \bigl\{\bigl(2+(1-c_1)\beta'\bigr)|\del\bm{\pi}|^2 
 - (3 - \gamma_m)m_\pi^2|\bm{\pi}|^2\bigr\}\sigma/(2F_\sigma)\,.
 \label{eqn:gspipi}
\end{equation}
Note the key feature of (\ref{eqn:gspipi}): it is mostly derivative so it has a 
small effect on $\pi\pi$-scattering in the $SU(2)_L\times SU(2)_R$ limit 
where $\del = O(m_\pi)$ and $m_s$ (and thus $m_\sigma^2$) remains 
fixed.  The vertex for an on-shell amplitude for $\sigma\to\pi\pi$ is readily 
obtained,
\begin{equation}
g_{\sigma\pi\pi} = -\bigl(2+(1-c_1)\beta'\bigr)m_\sigma^2/(2F_\sigma) 
+ O(m_\pi^2)\,,
\end{equation}
and generates part of the broad width for $\sigma$
\begin{equation}
      \Gamma_{\sigma\pi\pi} \approx \frac{|g_{\sigma\pi\pi}|^2}{16\pi m_\sigma} 
      \sim \frac{m_\sigma^3}{16\pi F_\sigma^2} \sim  250 \mbox{ MeV} \,.
\end{equation}
Note that $\Gamma_{\sigma\pi\pi} \sim O(m_\sigma^3)$ and thus NLO in 
the chiral-scale expansion.

\subsection{Weak Interactions and $K_S\to\pi\pi$}
\begin{figure}[t]
\center\includegraphics[scale=.7]{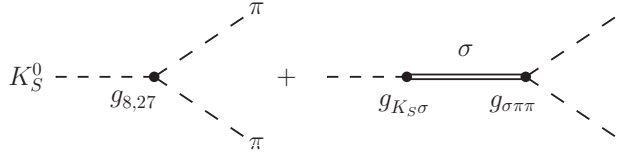}
\caption{Leading order diagrams in $\chpt_\sigma$ for $K_S\to\pi\pi$. 
The dominant $\sigma$-pole implies that the couplings 
$g_8$ and $g_{27}$ are allowed to be of similar magnitudes.} 
\label{fig:k_pipi}
\end{figure}

In $\chi$PT$_\sigma$, the anomalous dimension $\gamma_{mw}$ of 
$Q_M$ differs from that of $\mathcal{L}_{\mathrm{mass}}$.  Consequently, 
the chiral-scale Lagrangian includes a term 
$Q_M e^{(3-\gamma_{mw})\sigma/F_\sigma}$ whose $\sigma$ 
dependence cannot be eliminated by a chiral rotation. Instead, after vacuum alignment, 
the weak interactions are described in the LO of $\chpt_\sigma$ by
\begin{align}
\mathcal{L}^{\mathrm{align}}_{\mathrm{weak}}
 &= Q_{8}\sum_n g_{8n}e^{(2-\gamma_{8n})\sigma/F_\sigma}
 + g_{27}Q_{27}e^{(2-\gamma_{27})\sigma/F_\sigma}   \nonumber \\
 &+ Q_{mw}\bigl\{e^{(3-\gamma_{mw})\sigma/F_\sigma} 
 - e^{(3-\gamma_{m})\sigma/F_\sigma}\bigr\} + \mathrm{h.c.}\,,
\end{align}
noting that $Q_8$ represents quark-gluon operators with differing
dimensions at $\alpha_\mathrm{IR}$. As a result, $K_{S}$ and $\sigma$ 
mix through the interaction $\mathcal{L}_{K\sigma} = g^{}_{K\sigma}K_{S}^{0}\sigma$, 
where the effective coupling
\begin{equation}
g^{}_{K\sigma} =  (\gamma_m-\gamma_{mw})\Re\mathfrak{e} 
\{(2m_K^2-m\pi^2)\bar{g}_M -m_\pi^2 g_M\}  F_{\pi}/2F_{\sigma} \,,
\end{equation}
produces a $\Delta I = 1/2$ amplitude $A_{\sigma\textrm{-pole}}$ 
(Fig.~\ref{fig:k_pipi}).
%
%%% Fig.~\ref{fig:k_pipi}
The value of this coupling can be fixed by data on $\gamma\gamma\rightarrow\pi^{0}\pi^{0}$ 
and $K_{S}^{0}\rightarrow\gamma\gamma$, with the result
\begin{equation}
|g^{}_{K\sigma}| \approx 4.4\times 10^{3}\,\mathrm{keV}^{2}
\label{gksig} \end{equation}
accurate to about 30\% precision.  Subject to the uncertainty on $F_\sigma$, we 
estimate
\begin{equation}
\left|A_{\sigma\textrm{-pole}}\right|
\approx 0.34\,\mathrm{keV}\,,
\end{equation}
thereby accounting for the large $I=0\ \pi\pi$ amplitude $A_0$ in (\ref{eqn:ratio}).  
It follows that $g_8$ and $g_{27}$ are allowed to have similar magnitudes 
(with the latter fixed precisely by $K^+\to\pi^+\pi^0$) and thus the octet 
dominance hypothesis is no longer necessary to explain the $\Delta I =1/2$ 
rule.

\subsection{Bonus: Drell-Yan Ratio at $\alphir$}
Another welcome feature of $\chpt_\sigma$ is that is makes a 
{\it nonperturbative} prediction for the Drell-Yan ratio 
\begin{equation}
R = \frac{\sigma(e^{+}e^{-}\rightarrow\mathrm{hadrons})}%
{\sigma(e^{+}e^{-}\rightarrow\mu^{+}\mu^{-})}\,,
\end{equation}
at $\alpha_s=\alphir$.  This is so because the electromagnetic trace anomaly \cite{RJC72,Ell72}
\begin{equation}
\theta^{\mu}_{\mu} = \left.\theta^{\mu}_{\mu}\right|_{\mathrm{str.}} 
+ \frac{\alpha}{6\pi}R F_{\mu\nu} F^{\mu\nu}\,,
\end{equation}
implies an effective $\sigma\gamma\gamma$ coupling
\begin{equation}
\mathcal{L}_{\sigma\gamma\gamma} 
= \tfrac{1}{2} g_{\sigma\gamma\gamma} \sigma F_{\mu\nu}F^{\mu\nu}\,.
\label{eqn:L_sig2gam}
\end{equation}
Here $F_{\mu\nu}$ and $\alpha\simeq 1/137$ are the electromagnetic field strength and 
fine-structure constant.  Through direct calculation, (\ref{eqn:L_sig2gam}) can be 
shown to be \cite{CT12}
\begin{equation}
g_{\sigma\gamma\gamma} = \frac{(\rir - \tfrac{1}{2})\alpha}{3\pi F_\sigma}\,.
\end{equation}
Dispersive analyses \cite{Pen06,Pen07} of $\gamma\gamma\to\pi^0\pi^0$ indicate that 
the residue of the $f_0$ pole can be unambiguously extracted from data.  
Within the uncertainty in the value of $\fsig$, we use the updated value \cite{Oll08} 
$\Gamma_{f_0\gamma\gamma} = (1.98\pm 0.3)\,\mathrm{keV}$ to obtain 
\begin{equation}
\rir \approx 5\,.
\end{equation}

\section{Concluding Remarks}
We have seen that an infrared fixed point in the three flavor $\beta$-function 
of QCD leads to an extended NG sector $\{\pi,K,\eta,\sigma\}$, where the 
$f_0(500)$ resonance is identified with $\sigma$ as the dilaton of spontaneously 
broken scale symmetry.  Despite a seemingly drastic change to the accepted 
low-energy structure of QCD, the resulting chiral-scale perturbation theory 
$\chpt_\sigma$ is rather conservative: $f_0$ pole terms are promoted to 
leading order in $\chpt_\sigma$ (thereby evading $\chpt_3$'s convergence 
problems in the $0^{++}$ channel), yet the successful leading 
order predictions of $\chpt_3$ are preserved.  While our key result is a simple 
explanation for the $\Delta I=1/2$ rule in $K$-decays, $\chpt_\sigma$ is 
certainly a general framework and it would be interesting to examine the 
consequences of the effective theory for other well studied conundrums 
such as $CP$-violation, rare kaon decays, and $\eta\to 3\pi$.

\section*{Acknowledgements}
L.~C.~T.\ thanks the organizers and C.~Dominguez in particular for a most 
pleasant workshop and for the opportunity to present this work.  L.~C.~T.\  
has benefited from discussions with Profs.\ H.~Leutwyler, M.~Jamin, and 
P.~Minkowski, and thanks them for their useful comments.

%%% REFERENCES

\end{document}